\documentclass[12pt]{article}
\begin{document}
\begin{titlepage}
\begin{flushright}
hep-th/0302127\\
TIT/HEP-492\\
February, 2003
\end{flushright}
\vspace{0.5cm}
\begin{center}
{\Large \bf 
Penrose Limit and Enhan\c{c}on Geometry}
\lineskip .75em
\vskip2.5cm

{\large Katsushi Ito}\footnote{E-mail address:\ \ 
{\tt ito@th.phys.titech.ac.jp}} \ \ and \ \
{\large Yasuhiro Sekino}\footnote{E-mail address:\ \ 
{\tt sekino@th.phys.titech.ac.jp}}
\vskip 2.5em

{\large\it Department of Physics\\
Tokyo Institute of Technology\\
Tokyo, 152-8551, Japan}  
\vskip 4.5em
\end{center}

\begin{abstract}
We study superstring theories on the Penrose limit of 
the enhan\c{c}on geometry realized by 
the $D(p+4)$-branes wrapped on a $K3$ surface.
We first examine the null geodesics with fixed radius 
in general brane backgrounds, which  give solvable superstring 
theories with constant masses. In most cases, the superstring 
theories contain negative mass-squared.
We clarify a condition that the world-sheet free fields 
have positive mass-squared.
We then apply this condition to the enhan\c{c}on geometry and 
find that the null geodesics with fixed radius exist
only for $p=0$ case.
They define the superstring theories with positive
mass-squared.
For $p>0$ case, we show that there is no null geodesic with fixed
radius.
We also discuss the decoupling limit
which gives the dual geometry of super Yang-Mills theory
with 8 supercharges. We discuss the $K3$-volume 
dependence of the superstring spectrum.
\end{abstract}
\end{titlepage}
\baselineskip=0.7cm
\section{Introduction}

The work of Berenstein, Maldacena and Nastase 
(BMN) \cite{BeMaNa}
made a major step for understanding the
holographic correspondence of string theory
and gauge theories.
Superstring theory on the maximally supersymmetric
plane wave~\cite{BlFiHuPa}, which is given by the Penrose 
limit~\cite{Pe, BlFiPa} of $AdS_{5}\times S^{5}$, 
is exactly solvable~\cite{Me}. 
Based on this result and 
the AdS/CFT correspondence~\cite{Ma},
BMN identified the string states on the 
plane wave background as certain
sectors of the $D=4$, ${\cal N}=4$ super Yang-Mills theory.
Supersymmetric gauge theories with conformal invariance but
less supersymmetry
has been also investigated 
extensively~\cite{BeGaMaNaNa}-\cite{NiPr}.

It is an interesting problem to explore 
holographic correspondences for 
non-conformal gauge theories \cite{IzMaSoYa, JeYo,
AhBeKuSe}
using string theories on brane backgrounds. 
In a previous paper~\cite{FuItSe}, Fuji and the present 
authors have studied string theories on the Penrose limits of the
brane solutions including the D$p$-brane, $(p,q)$ fivebrane and
$(p,q)$ string solutions. We have solved exactly
the equations of motion of bosonic strings
on certain backgrounds, which have time-dependent masses.
Aspects of the quantization of strings
on time-dependent plane wave backgrounds are
discussed in refs.~\cite{FuItSe, PaRuTs}.

String theories on the Penrose limits
of brane solutions and the possible connections to 
the gauge theories have been studied in the works such as
ref.~\cite{GiPaSo,AlKu} for the D$p$-branes, 
and refs.~\cite{GoOo, HuRaVe, OzSa} for the
fivebrane solutions. 
The Pilch-Warner solution~\cite{PiWa1}, which provides the
dual of the RG flow from the ${\cal N}=4$ super
Yang-Mills theory to the superconformal 
${\cal N}=1$ theory of Leigh and Strassler~\cite{LeSt}, 
have been studied in the Penrose limit~\cite{GiPaSo,CoHaKeWa}. 
String theories were analyzed near the region of the
geometry corresponding to the IR fixed point.
Recently,  Gimon {\it et.al.} \cite{GiPaSoSt} 
studied the Maldacena-N\'{u}\~{n}ez~\cite{MaNu} and 
the Klebanov-Strassler~\cite{KlSt} solutions, 
which are dual to ${\cal N}=1$ theories. 
They considered the Penrose limit along a special class 
of null geodesics which give solvable string theories.
It was argued that the string spectrum represent 
hadrons in the IR limit of the gauge theories.

In this paper, we study geometries
which are dual to super Yang-Mills theories  
with 8 supercharges.
The brane system corresponds to the 
D-branes wrapped on $K3$, which is described by
the `enhan\c{c}on' geometry \cite{JoPePo}.
When D$(p+4)$-branes are wrapped on $K3$, 
the curvature of the $K3$ induces negative D$p$-brane charges.
In the supergravity, they are effectively described by a
geometry which is similar to the 
D$(p+4)$-$\overline{{\rm D}p}$-brane system. 
This geometry is valid outside the 
`enhan\c{c}on radius'. Inside the radius,
we have flat space.  
It is proposed that this enhan\c{c}on geometry
is dual to the $(p+1)$-dimensional 
super Yang-Mills theory \cite{JoPePo}. 
As a result of the $K3$ compactification, the gauge theory
has half the maximal supersymmetry, and contain the massless
fields which are the same as 
the $D=4$, ${\cal N}$=2 super Yang-Mills theory 
with no hypermultiplets.

We study superstring theory on the Penrose limit of 
enhan\c{c}on geometry. 
We investigate the general null geodesics of the 
enhan\c{c}on geometry for D$(p+4)$-branes wrapped on $K3$.
Our particular interest is the special null geodesics 
such that the radial transverse coordinate is fixed.
The Penrose limit along such geodesics 
gives plane waves on which the light-cone
string theory has constant masses.

In recent studies~\cite{FuItSe, GiPaSo}, 
it has been noted that 
many plane wave backgrounds which are obtained 
by the Penrose limits yield
string theories which have negative mass-squared. 
Although stability analysis based on supergravity suggests
the background is stable \cite{MaPa}, it is difficult to make 
sense of those string theories which have 
instability on the world-sheet. 
The ground state of the string would be a stretched
configuration due to the tidal force of the brane
background~\cite{GiPaSo}. 
In this paper, we add an
example of string theory where all the mass-squared 
are non-negative.  In such a case, 
we expect that the string spectrum and their interpretation
in terms of the gauge theory are clearly understood.

On the enhan\c{c}on geometry, 
we find that the null geodesics with constant radius
exist only for the $p=0$ case (D4-branes wrapped on $K3$).
Moreover, in this case, we obtain a superstring theory 
whose mass-squared are all non-negative. 
We will calculate the superstring spectrum 
based on the light-cone Hamiltonian.
By taking the decoupling limit, we can discuss the 
correspondence with the gauge theory. 
In the present work, however,
we do not discuss correspondence between 
string states and gauge-theory operators.
We argue that our results of the string theory should
give information on the gauge theory 
in the large $N$ limit with large effective 't Hooft coupling.
We also note a qualitative behavior of the string spectrum
as the volume of $K3$ varies. 
For $p=1,2$ cases, we show that 
null geodesic with constant radius does not exist.

This paper is organized as follows. 
We first review the enhan\c{c}on geometry in section~2. 
In section~3, we discuss 
the null geodesic with constant radius 
in general brane backgrounds
and the Penrose limit along such a geodesic. 
In section~4, we examine null geodesics in 
the enhan\c{c}on geometry associated with
D$(p+4)$-branes wrapped on $K3$. 
We take the Penrose limit along the geodesic with
constant radius for  the $p=0$ case.
In section~5, we study
the spectrum of the superstring theory in this Penrose limit.
Discussion on the dual gauge theory is given in section~6.
Conclusions and discussion are in section~7.

We include an Appendix, in which we apply the 
general discussion 
on the null geodesics with constant radius
to other brane solutions.
We analyze D$p$-brane, $(p,q)$ fivebrane, 
$(p,q)$ string solutions, which have been treated 
in the previous paper~\cite{FuItSe}.
The conditions for the existence of the null geodesics 
with constant radius on these backgrounds are 
carefully reexamined.

\section{Enhan\c{c}on geometry}

We begin with  a brief review on the
enhan\c{c}on geometry. 
When D$(p+4)$-branes are wrapped on the whole $K3$, 
the curvature of the $K3$ induces one unit of 
negative D$p$-brane charge per one D$(p+4)$-brane. 
The supergravity solution which describes this system
in the string frame is given 
as follows~\cite{JoPePo}
\begin{eqnarray}
ds^{2}&=&Z_{p}^{-1/2}Z_{p+4}^{-1/2}(-dt^{2}+dx_{a}^{2})
+Z_{p}^{1/2}Z_{p+4}^{1/2}(dr^{2}+r^{2}d\Omega_{4-p}^{2})
\nonumber\\
&&+V^{1/2}Z_{p}^{1/2}Z_{p+4}^{-1/2}ds^{2}_{K3},
\label{eq:enhancon}\\
A_{p+1}&=&{1\over g_{s}Z_{p}}dx^{0}\wedge d{\rm vol}(R^{p}),
\label{eq:enhA1}\\
A_{p+5}&=&{V\over g_{s}Z_{p+4}}dx^{0}\wedge d{\rm vol}(R^{p})\wedge
d{\rm vol}(K3),
\label{eq:enhA5}\\
e^{\phi}&=&g_{s}Z^{(3-p)/4}_{p}Z^{-(p+1)/4}_{p+4},
\label{eq:dilaton}\\
Z_{p}&=&1-{\tilde{r}_{p}^{3-p}\over r^{3-p}},\quad
Z_{p+4}=1+{r_{p+4}^{3-p}\over r^{3-p}},
\label{eq:Z}
\end{eqnarray}
where $x^{a}$ $(a=1,\ldots, p)$ are the coordinates along 
the brane other than those wrapped on the $K3$. $r$ is the
radial coordinate in the transverse $(5-p)$-dimensional space.
$A_p$ denotes the RR $p$-form, and $\phi$ is the dilaton.
The string coupling at the infinity is given by $g_s$.
$ds^{2}_{K3}$ is the metric of the $K3$ with unit volume, 
$d{\rm vol}(K3)$ is the 
corresponding volume form, and the constant $V$ is 
the volume of $K3$ measured at the infinity.  
The parameters $\tilde{r}_{p}$ and $r_{p+4}$, which are
positive, are given by
\begin{equation} 
r_{p+4}^{3-p}=(2\sqrt{\pi})^{1-p}\Gamma \left(
{3-p\over 2}\right)
g_{s}N\alpha'{}^{3-p\over 2},\quad
\tilde{r}_{p}^{3-p}=
{V_{*}\over V}r_{p+4}^{3-p},
\end{equation}
where $V_{*}\equiv (2\pi)^{4}\alpha'{}^{2}$
and $N$ is the number of the D$(p+4)$-branes.
We will consider the cases of $p=0,1,2$. 

As explained in ref.~\cite{JoPePo},
this geometry is valid outside the `enhan\c{c}on 
radius' $r_{e}$. The D$(p+4)$-branes form a shell 
at the enhan\c{c}on radius,
and the geometry inside is replaced with the flat space.
A way to find $r_{e}$ is to consider a 
probe D$(p+4)$-brane (also wrapped on $K3$). 
The enhan\c{c}on radius is determined as 
the radius at which the tension
of the probe brane vanishes. 
Following the arguments in refs.~\cite{JoPePo, JoMyPeRo}, 
we have
\begin{equation}
r_{e}^{3-p}={2V\over V-V_{*}}\tilde{r}_{p}^{3-p}.
\end{equation}
The metric (\ref{eq:enhancon}) has
a naked singularity called the
`repulson singularity' at $r=\tilde{r}_{p}$,
where repulsive force on a test particle diverges, 
but it is removed since 
the singularity is in the unphysical 
region $\tilde{r}_{p}<r_{e}$.

It is proposed that the decoupling limit of 
this geometry is dual to super Yang-Mills theory  with 
8 supercharges \cite{JoPePo}. The decoupling
limit is given by replacing $Z_{p}$ and $Z_{p+4}$ 
in (\ref{eq:Z}) with
\begin{equation}
Z_{p}=1-{\tilde{r}_{p}^{3-p}\over r^{3-p}},\quad
Z_{p+4}={r_{p+4}^{3-p}\over r^{3-p}}.
\label{eq:decoupling}
\end{equation}
This is obtained  from the full geometry by 
taking $r\ll r_{p+4}$ with $r \sim \tilde{r}_{p}$.
In other words, this corresponds to a 
`near-shell limit' with $V/V_{*}$ taken large. 
Also note that the enhan\c{c}on radius for the metric 
in the decoupling limit is  given by
$r_{e}^{3-p}=2\tilde{r}_{p}^{3-p}$.

\section{Null geodesics with constant radius in general 
brane backgrounds}

The main purpose of this paper is to study enhan\c{c}on
geometry in the Penrose limit along null geodesics which 
stay at constant radius. 
Before beginning to study the enhan\c{c}on geometry,
we shall discuss some general properties of the null 
geodesics with constant radius. 
Assuming that the metric takes the form 
of generic brane solutions,
we derive the condition for the existence of the null 
geodesic with constant radius. We then obtain the formula
for the Penrose limit along such a geodesic.

\subsection{Geodesic equations}

Let us consider generic brane solutions in $D$ 
spacetime dimensions,
which are static, rotationally symmetric in the transverse space, 
and homogeneous in the spatial directions 
along the brane:
\begin{equation}
ds^{2}=-A^{2} dt^{2} + \tilde{A}^{2} dx_{a}^{2}
+B^{2}dr^{2} +\tilde{B}^{2}r^{2}d\Omega_{D-p-2}^{2}.
\label{eq:pbr1}
\end{equation}
where $x^{a} (a=1,\ldots, p)$ are the coordinates 
along the branes.
$A$, $\tilde{A}$, $B$ and $\tilde{B}$ are functions
of the transverse radial distance $r$. 
Here, $d\Omega_{D-p-2}^{2}$
is the metric of the unit sphere $S^{D-p-2}$. 
We will use the following parametrization of $S^{D-p-2}$:
\begin{equation}
d\Omega_{D-p-2}^{2}=\cos^2 \theta d\psi^2
+d\theta^2 +\sin^2\theta d\Omega_{D-p-4}^{2}.
\label{eq:sphere}
\end{equation}

We consider null geodesics in the $(t,r,\psi)$ space.
A null geodesic is given by the trajectory of 
a test particle with the action
\begin{eqnarray}
S&=&{1\over 2}\int d\tau  g_{\mu\nu}\dot{x}^{\mu}
\dot{x}^{\nu} \nonumber\\
&=&{1\over 2} \int d\tau \left( -A^{2}\dot{t}^{2}
+B^{2}\dot{r}^{2}+\tilde{B}^{2}r^{2}\dot{\psi}^{2}\right)
\end{eqnarray}
and the massless constraint 
\begin{equation}
g_{\mu\nu} \dot{x}^{\mu}\dot{x}^{\nu}=
-A^{2}\dot{t}^{2}+B^{2}\dot{r}^{2}
+\tilde{B}^{2}r^{2}\dot{\psi}^{2}
=0
\end{equation}
where $\cdot\equiv d/d\tau$. 

From the equations of motion for $t$ and $\psi$, 
we see that the energy $E\equiv A^{2}\dot{t}$ and 
the angular momentum
$J\equiv \tilde{B}^{2}r^{2}\dot{\psi}$ are conserved.
We rescale $\tau \to \tau/E$ and 
normalize the energy to one. Then, we have
\begin{equation}
\dot{t}={1\over A^{2}},\quad 
\dot{\psi}={\ell\over \tilde{B}^{2}r^{2}}, 
\end{equation}
where $\ell\equiv J/E$.
The massless constraint becomes
\begin{equation}
-{1\over A^{2}}+B^{2}\dot{r}^{2}
+{\ell^{2}\over \tilde{B}^{2}r^{2}}=0.
\label{eq:massless}
\end{equation}
The velocity in the radial direction $\dot{r}$ is
determined as
\begin{equation}
\dot{r}=\pm \sqrt{{1\over B^{2}A^{2}}
-{\ell^{2}\over B^{2}\tilde{B}^{2}r^{2}}}.
\label{eq:dotr}
\end{equation}
Since $\dot{r}^{2}\ge 0$, (\ref{eq:massless}) gives
the region of $r$, which is determined by
\begin{equation}
{1\over B^{2} A^{2}}\ge {\ell^{2}\over B^{2}\tilde{B}^{2}r^{2}}.
\label{eq:geodcond}
\end{equation}

The points where the equality in (\ref{eq:geodcond}) holds
are generically the turning points of the geodesic.
The direction to which the geodesic turns is determined
by the second time-derivative $\ddot{r}$ at that point.
{}From the equation of motion for $r$:
\begin{equation}
-2 B^{2}\ddot{r}-(A^{2})'\dot{t}^{2}-(B^{2})'\dot{r}^{2}
+(\tilde{B}^{2}r^{2})'\dot{\psi}^{2}=0,
\end{equation}
we obtain
\begin{equation}
\ddot{r}={1\over 2B^{2}}\left(
-{(A^{2})'\over A^{4}} +{(\tilde{B}^{2}r^{2})' \ell^{2}\over
\tilde{B}^{4}r^{4}} -(B^{2})'\dot{r}^{2}\right)
\label{eq:ddotr}
\end{equation}
where $'\equiv d/dr$.

We may consider a geodesic which stays at a fixed
radius when $r$ and $\ell$ are adjusted such that 
$\dot{r}=\ddot{r}=0$ are satisfied. 
From (\ref{eq:dotr}) and (\ref{eq:ddotr}), we find
that the fixed radius conditions are given by
\begin{equation}
{1\over B^{2}}\left( {1\over A^{2}}
-{\ell^{2}\over \tilde{B}^{2}r^{2}}
\right)=0,\qquad
{1\over B^{2}}\left( {(A^{2})'\over A^{4}}
-{(\tilde{B}^{2}r^{2})' \ell^{2}\over
\tilde{B}^{4}r^{4}}\right) =0.
\label{eq:constcond1}
\end{equation}
It is useful to define the function
\begin{equation}
F(r)\equiv -A^{2}+{\tilde{B}^{2}r^{2}\over \ell^{2}}.
\label{eq:defFr}
\end{equation}
The conditions (\ref{eq:constcond1}) are equivalent to
\begin{equation}
F(r)=0, \qquad F'(r)=0,
\label{eq:constcond}
\end{equation}
if $\ell^{2}/(A^{2}B^{2}\tilde{B}^{2}r^{2})\neq 0$.

\subsection{Penrose limit along a null geodesic with constant radius}
When a null geodesic  stays at a constant radius,
we can take the Penrose limit along it in the 
following way.
The coordinates of the geodesic are parametrized by
\begin{equation}
t={1\over A^{2}_{*}}\tau,\quad 
\psi={\ell_{*}\over \tilde{B}_{*}^{2}r_{*}^{2}}\tau,
\quad r=r_{*}
\end{equation}
where $r_{*}$ and $\ell_{*}$ are the solutions
of  (\ref{eq:constcond}).
$A_{*}$ stands for  $A(r_{*})$ etc.

We introduce the light-cone coordinates
\begin{equation}
x^{\pm}\equiv {1\over 2}\left( A^{2}_{*} t
\pm {\tilde{B}^{2}_{*}r^{2}_{*}\over \ell_{*}}\psi \right),
\label{eq:coordtr}
\end{equation}
and rewrite the metric (\ref{eq:pbr1}) of the
brane solution as
\begin{eqnarray}
ds^{2}&=&-{A^{2}\over A^{4}_{*}}(dx^{+}+dx^{-})^{2}
+{\tilde{B}^{2}r^{2}\cos^{2}\theta\ell^{2}_{*}
\over \tilde{B}_{*}^{4}r^{4}_{*}}(dx^{+}-dx^{-})^{2}
\nonumber\\
&&+\tilde{A}^{2}dx_{a}^{2}+B^{2}dr^{2}
+\tilde{B}^{2}r^{2}(d\theta^{2}+\sin^{2}\theta d\Omega^{2}_{6-p}).
\label{eq:pbrlc}
\end{eqnarray}
The Penrose limit is given by the following combination of
the coordinate transformation and the rescaling of the metric,
\begin{eqnarray}
&&x^{+}\to x^{+},\quad x^{-}\to \Omega^{2}x^{-},
\quad z\to \Omega z,\quad 
\theta \to\Omega\theta,\quad x^{a}\to \Omega x^{a},\nonumber\\
&&ds^{2}\to \Omega^{-2}ds^{2},
\end{eqnarray}
in the limit $\Omega\to 0$.
Here $z$ is a shifted radial coordinate 
$z\equiv r-r_{*}$.

The divergent terms in (\ref{eq:pbrlc}) cancel due to
(\ref{eq:constcond1}). 
The parts which remain finite
take the plane wave form. After rescaling the coordinates
by suitable constants, we obtain\footnote{%
In this paper we have changed the sign
of the definition of $m^{2}_{i}$ in (\ref{eq:planewave})
from the one in the previous paper~\cite{FuItSe}.}
\begin{eqnarray}
ds^{2}&=&2dx^{+}dx^{-} -\left(m^{2}_{x}\sum_{a=1}^{p}x^{2}_{a} 
+m^{2}_{y}\sum_{l=1}^{D-p-3}y^{2}_{l}
+m^{2}_{z}z^{2}\right)(dx^{+})^{2}\nonumber\\
&&\quad +\sum_{a=1}^{p}dx_{a}^{2}
+\sum_{l=1}^{D-p-3}dy_{l}^{2}+dz^{2}.
\label{eq:planewave}
\end{eqnarray}
The coefficients $m_{i}$ are constant and are given by
\begin{equation}
m^{2}_{x}=0,\quad
m^{2}_{y}={\ell^{2}_{*}\over \tilde{B}^{4}_{*}r^{4}_{*}},\quad
m^{2}_{z}=-{1\over 2}{1\over B^{2}_{*}A^{4}_{*}}F''(r)_{*}
\label{eq:constmass}
\end{equation}
where $F''(r)_{*}\equiv F''(r)|_{r=r_{*}}$, and 
$F(r)$ is defined in (\ref{eq:defFr}).
The coordinates $y_{l}$ ($l=1,\ldots,D-p-3$) are
defined by
\begin{equation}
\sum_{\ell=1}^{D-p-3}
d y^{2}_{l}= d\theta^{2} +\theta^{2}d \Omega^{2}_{D-p-4}.
\end{equation}

As we will see in section 5, $m^{2}_{i}$ are the mass-squared 
(times a positive constant) for the bosonic modes $X^{i}$ 
of the string in the light-cone gauge on this background.
Note that the sign of $m^{2}_{y}$ is always positive,
but the sign of $m^{2}_{z}$ is given by 
the sign of $-F''(r)_{*}$. 
In the present work, we will study string theory on 
the plane wave where
$m^{2}_{i}$ are all non-negative. When some of the
$m^{2}_{i}$ become negative, there will be an instability 
on the world-sheet. Note that this does not necessarily
mean an instability of the spacetime (\ref{eq:planewave}).  
In fact, the analyses of the linearized fluctuations
around some plane waves with negative $m^{2}_{i}$ suggests
that the spacetime is stable \cite{MaPa}.
This instability on the world-sheet suggests that the ground state
of the string is not point-like, but is rather a stretched 
configuration, due to the tidal force, as mentioned
in ref.~\cite{GiPaSo}.
Semi-classical analysis of strings around expanded 
configurations were performed in refs.~\cite{GuKlPo} 
for the $AdS_{5}\times S^{5}$ background.
Study of string theories on various brane backgrounds
along this line would be interesting, but it is 
beyond the scope of the present paper. 

In Appendix, we apply the general discussion 
in this section to various brane backgrounds.
We examine the conditions for the fixed radius null
geodesics for the D$p$-brane, $(p,q)$ fivebrane,
$(p,q)$ string solutions. We will find several 
examples of the null geodesics with fixed radius. 
We will also see that the resulting string theories
have non-negative mass-squared only for 
the near-horizon limits of the fivebrane solutions.

\section{Penrose limit of enhan\c{c}on geometry}
We now study the Penrose limit of enhan\c{c}on
geometry. 
In the first subsection, we will discuss 
possible null geodesics 
in the enhan\c{c}on geometries associated with D$(p+4)$-branes.
In particular, we look for the null geodesics which 
stay at constant radius, by
using the formulas in the previous section.
We will show that such geodesics exist for the $p=0$ case. 
In the second subsection, we will obtain the Penrose
limit of the $p=0$ enhan\c{c}on geometry
along the geodesic.

\subsection{Possible types of the null geodesics}

We consider null geodesics 
in the $(t,r,\psi)$ space in the enhan\c{c}on geometry, 
where $r$ is the radial coordinate
and  $\psi$ is one of the angular coordinates on the
$S^{4-p}$ in the transverse space. We assume that
the geodesic is fixed in the $K3$ or in the 
directions $x^{a}$ along the $p$-brane.
We may use the formulas in the last section by setting
$D=6$, $A^{2}=Z_{p}^{-1/2}Z_{p+4}^{-1/2}$ and 
$B^{2}=\tilde{B}^{2}=Z_{p}^{1/2}Z_{p+4}^{1/2}$.
For our background, it is convenient to define 
\begin{eqnarray}
f(r)&\equiv&
r^{4-2p}\ell^{2}Z^{1/2}_{p}Z^{1/2}_{p+4}F(r)\nonumber\\
&=&r^{6-2p}+(r_{p+4}^{3-p}-\tilde{r}_{p}^{3-p})r^{3-p}
-\ell^{2}r^{4-2p}-\tilde{r}_{p}^{3-p}r_{p+4}^{3-p}.
\label{eq:enhfr}
\end{eqnarray}
The condition (\ref{eq:geodcond}) for the allowed region
of a geodesic is equivalent to $f(r)\ge 0$.
Moreover, the condition for the fixed radius 
$\dot{r}=\ddot{r}=0$ is equivalent to $f(r)=f'(r)=0$,
and the sign of $-f''(r)$ evaluated at the fixed radius
is the same as the sign of
$m^{2}_{z}$.

Firstly we examine what kind of null geodesics are possible for
the cases of $p=1$ and $p=2$. We can see from (\ref{eq:enhfr}),
that we have $f(r)>0$ for sufficiently large $r$,
and that $f(r)=0$ has one solution $r=r_{T}$. 
If the parameter $\ell$ is chosen such that 
$r_{T}$ is in the physically sensible region 
($r_{T}\ge r_{e}$), $r_{T}$ is a turning point 
where a geodesic which comes from $r=\infty$ 
turns outward. On the other hand, if 
$r_{T}<r_{e}$, the geodesic from the infinity
reaches the enhan\c{c}on radius, and it would pass
through the flat region ($r<r_{e}$) and then goes back to 
infinity.
We can easily see that $f(r)=f'(r)=0$
do not have a solution, and there is not a geodesic 
which stay at constant radius for $p=1,2$.

For the $p=0$ case, on the other hand,
null geodesics with fixed radius 
is possible. The conditions for the fixed radius 
for $p=0$ become
\begin{eqnarray}
f(r)&=&r^{6}-\ell^{2}r^{4}+(r_{4}^{3}-\tilde{r}_{0}^{3})r^{3}
-\tilde{r}_{0}^{3}r_{4}^{3}=0,
\label{eq:fr0}\\
f'(r)&=&6r^{5}-4\ell^{2}r^{3}
+3(r_{4}^{3}-\tilde{r}_{0}^{3})r^{2}=0.
\label{eq:frp0}
\end{eqnarray}
By eliminating $\ell^{2}$ from (\ref{eq:fr0}) and (\ref{eq:frp0}),
we obtain the equation which determines the fixed radius
\[
2r^{6}-(r_{4}^{3}-\tilde{r}_{0}^{3})r^{3}
+4\tilde{r}_{0}^{3}r_{4}^{3}=0,
\]
which have two solutions
\begin{equation}
r^{3}_{*\pm}={1\over 4}\left(
x-1\pm\sqrt{(x-1)^{2}-32 x}\right)\tilde{r}_{0}^{3}.
\label{eq:rastpm}
\end{equation}
Here we have set $x\equiv V/V_{*}$, and used 
the relation $r_{4}=x\tilde{r}_{0}$.
The formal solutions in (\ref{eq:rastpm})
are both positive  when 
$x-1>0$ and $(x-1)^{2}-32x>0$, 
which is equivalent to
\begin{equation}
x\ge 17+12\sqrt{2}.
\label{eq:xineq}
\end{equation}
The two positive solutions $r_{*\pm}$
in this case
are both in the physical region, {\it i.e.} outside
the enhan\c{c}on radius $r_{e}$. In fact, 
the difference $r_{*-}^{3}-r_{e}^{3}$ is given by
\[
r^{3}_{*-}-r^{3}_{e}={1\over 4}
\left\{ x-1 -{8x\over x-1}-\sqrt{(x-1)^{2}-32x} 
\right\}\tilde{r}_{0}^{3}.
\]
When (\ref{eq:xineq}) holds, 
the r.h.s. is positive since
\[
\{ (x-1)^{2}-8x \}^{2}-(x-1)^{2}\{ (x-1)^{2}-32x \}
=16x(x+1)^{2} >0.
\]
Thus, $r_{*-}$ and $r_{*+}$ are greater than $r_{e}$.
If (\ref{eq:xineq}) is not satisfied, 
(\ref{eq:fr0}) and (\ref{eq:frp0}) do not have positive
solution for any $\ell$.

%

If the $K3$ volume $V$ is large enough to 
satisfy (\ref{eq:xineq}), we may consider the null geodesic 
which is fixed at $r=r_{*+}$ or at $r=r_{*-}$. 
The parameter $\ell$ is fixed to $\ell_{*+}$ or to $\ell_{*-}$, 
correspondingly, by the condition $f'(r)=0$: 
\begin{equation}
\ell_{*\pm}^{2}={3\over 2}r_{*\pm}^{2}
+{3\over 4r_{*\pm}}(r_{4}^{3}-\tilde{r}_{0}^{3}).
\label{eq:last}
\end{equation}
Evaluating $f''(r)$ with $r=r_{*\pm}$, $\ell=\ell_{*\pm}$,
we obtain
\[
f''(r)=\pm 3r_{*\pm}\sqrt{(r_{4}^{3}-\tilde{r}_{0}^{3})^{2}
-32\tilde{r}_{0}^{3}r_{4}^{3}}.
\]
That is, we have $f''(r)<0$ at $r=r_{*-}$, 
but $f''(r)>0$ at $r=r_{*+}$. 

We may analyze the null geodesics in 
the enhan\c{c}on geometries in the 
decoupling limit (\ref{eq:decoupling})
by replacing the function $f(r)$ in 
(\ref{eq:enhfr}) with
\begin{equation}
f(r)=-\ell^{2}r^{4-2p}+r_{p+4}^{3-p}r^{3-p}
-\tilde{r}_{p}^{3-p}r_{p+4}^{3-p}.
\label{eq:enhfrdec}
\end{equation}
For $p=0$, the condition for 
the constant radius $f(r)=f'(r)=0$ has one
solution:
\begin{equation}
r^{3}_{*d}=4\tilde{r}^{3}_{0},\quad
\ell^{2}_{*d}={3\over 4}{r_{4}^{3}\over r_{*d}}.
\label{eq:rastdec}
\end{equation}
When evaluated with this solution, we have  
$f''(r)=-3r_{4}^{3}r_{*d}<0$.
This radius $r_{*d}$ is greater than the enhan\c{c}on
radius $r_{e}^{3}=2\tilde{r}_{0}^{3}$. 
Note that this $r_{*d}$ corresponds to $r_{*-}$ 
in the full geometry. Indeed, we may obtain (\ref{eq:rastdec})
by taking the $x\to\infty$ limit for $r_{*-}$ in
(\ref{eq:rastpm}). 
The larger solution $r_{*+}$ is 
scaled out in the decoupling limit.
For the $p=1,2$, we do not have a solution of 
$f(r)=f'(r)=0$ with $f(r)$ in (\ref{eq:enhfrdec}).

\subsection{Penrose limit of the $p=0$ enhan\c{c}on along the 
null geodesic with constant radius}

We take the Penrose limit along the null geodesic 
with constant radius which we have found for 
the $p=0$ enhan\c{c}on. We can readily obtain the
plane wave using the formula in section 3.

From (\ref{eq:planewave}) and (\ref{eq:constmass}),
we get
\begin{equation}
ds^{2}=2dx^{+}dx^{-}-(m^{2}_{y}y_{l}^{2}+m^{2}_{z}z^{2})(dx^{+})^{2}
+dy_{l}^{2}+dz^{2}+dw_{s}^{2}
\label{eq:planewaveenh}
\end{equation}
where 
coordinates $w_{s}$ 
$(s=5,6,7,8)$ are the ones along the $K3$.
In our case where the geodesic is fixed in the $K3$,
the $w_{s}$-part of the metric becomes flat. 
Note that we do not have $x^{a}$ (spatial directions along
the $p$-brane), since $p=0$.
The constant coefficients $m^{2}_{y}$ and $m^{2}_{z}$ are 
given by
\begin{equation}
m^{2}_{y}={\ell^{2}_{*}\over Z_{0*}Z_{4*}r^{4}_{*}},
\quad
m^{2}_{z}=-{1\over 2}{1\over \ell^{2}_{*}r^{4}_{*}}f''(r)_{*}.
\label{eq:masses}
\end{equation}
where $Z_{p*}$, $Z_{p+4*}$ mean 
$Z_{p}(r_{*})$, $Z_{p+4}(r_{*})$,
and $f''(r)_{*}\equiv f''(r)|_{r=r_{*}}$.
We denote  the solutions of $f(r)=f'(r)=0$,
which are given in (\ref{eq:rastpm}) and (\ref{eq:last})
or in (\ref{eq:rastdec}), 
by $r_{*}$ and $\ell_{*}$. 

If the Penrose limit is taken along $r_{*-}$ or 
$r_{*d}$, $m^{2}_{z}$ is positive, but
if the limit is taken along $r_{*+}$, $m^{2}_{z}$ is negative,
as we see from the signs of $-f''(r)_{*}$.
We will study string theory in the former cases, for the 
reason mentioned at the end of section~3. 
Substituting $r_{*}=r_{*-}$, $\ell_{*}=\ell_{*-}$ into
(\ref{eq:masses}), we obtain the explicit forms
of $m^{2}_{i}$ as follows: 
\begin{eqnarray}
m^{2}_{y}&=&{3\over 2}
{(x-1-h(x))(3x-3-h(x))\over (x-5-h(x))(5x-1-h(x))}
{1\over r_{*-}^{2}},
\nonumber\\ 
m^{2}_{z}&=& {4 h(x)\over (3x-3-h(x))}
{1\over r_{*-}^{2}},
\label{eq:massrm}
\end{eqnarray}
where 
\[
h(x)\equiv \sqrt{(x-1)^{2}-32x},
\]
and $r_{*-}$ is given in (\ref{eq:rastpm}). 
The plane wave for the geometry in the decoupling limit,
obtained by substituting $r_{*}=r_{*d}$, 
$\ell_{*}=\ell_{*d}$ in (\ref{eq:masses}),
is given by
\begin{equation}
m^{2}_{y}={1\over r_{*d}^{2}},\quad
m^{2}_{z}={2\over r_{*d}^{2}},
\label{eq:massdec}
\end{equation}
where $r_{*d}$ is in (\ref{eq:rastdec}).
Note that (\ref{eq:massdec}) can be obtained
from (\ref{eq:massrm}) 
by taking the $x\to \infty$ limit.

The RR fields in the Penrose limit are obtained 
by applying the coordinate transformations and the rescalings of
the coordinates mentioned in section 3.2, to
(\ref{eq:enhA1}), (\ref{eq:enhA5}):
\begin{eqnarray}
F_{2}&=&dA_{1}
={\ell_{*}\over g_{s}r_{*}}{1\over Z^{1/4}_{0*}Z^{1/4}_{4*}}
\left( {1\over Z_{0}}\right)'_{*}dx^{+}\wedge dz,\nonumber\\
F_{6}&=&dA_{5}
={\ell_{*}\over g_{s}r_{*}}{Z^{3/4}_{4*}\over Z^{5/4}_{0*}}
\left( {1\over Z_{4}}\right)'_{*}dx^{+}\wedge dw^{5}
\wedge dw^{6}\wedge dw^{7}\wedge dw^{8}\wedge dz.
\end{eqnarray}
Here, we have used the relation
\[
VZ_{0}Z_{4}^{-1}d{\rm vol}(K3)\to 
dw^{5}\wedge dw^
{6}\wedge dw^{7}\wedge dw^{8},
\]
which follows from the fact that
the $w$-part of the metric becomes flat in the
Penrose limit.
In the following, we will use the field strength $F_{4}$,
dual to $F_{6}$: 
\begin{equation}
F_{4}={\ell_{*}\over g_{s}r_{*}}{Z^{3/4}_{4*}\over Z^{5/4}_{0*}}
\left( {1\over Z_{4}}\right)'_{*}dx^{+}\wedge dy^{1}
\wedge dy^{2}\wedge dy^{3}.
\end{equation}

The explicit forms of the RR-field strengths (times
a dilaton factor) for $r_{*}=r_{*-}$ are given by
\begin{eqnarray}
e^{\phi}F_{+z}&=&-2^{3/2}\cdot 3^{3/2}
{(3x-3-h(x))^{1/2}(x-1-h(x))^{1/2}
\over (5x-1-h(x))^{1/2}(x-5-h(x))^{3/2}}
{1\over r_{*-}},
\nonumber\\
e^{\phi}F_{+123}&=&2^{3/2}\cdot 3^{3/2}
{ x(3x-3-h(x))^{1/2}(x-1-h(x))^{1/2}
\over (5x-1-h(x))^{3/2}(x-5-h(x))^{1/2}}
{1\over r_{*-}}.
\label{eq:fluxesrm}
\end{eqnarray}
For $r_{*}=r_{*d}$, we have
\begin{equation}
e^{\phi}F_{+z}
=-{1\over r_{*d}},\quad
e^{\phi}F_{+123}
= {3\over r_{*d}}.
\label{eq:fluxes}
\end{equation}

\section{Superstring spectrum on the plane wave} 

Having obtained the Penrose limit of the
$p=0$ enhan\c{c}on, we now study superstring theory
on the plane wave background. 
Our convention for the fermionic
part of the action follows that of 
Cveti\v{c} {\it et.al.}~\cite{CvLuPoSt}.
The covariant action of type IIA superstring 
up to the quadratic
order in fermions\footnote{It has been shown 
that the superstring action in the light-cone gauge is 
quadratic in fermions for fairly general
pp-wave backgrounds~\cite{MiMoSa}.}
is given by 
\begin{eqnarray}
S&=&{1\over 2\pi\alpha'}\int d\tau \int_{0}^{2\pi}
d\sigma\left({\cal L}_{b}+{\cal L}_{f}\right)
\nonumber\\
{\cal L}_{b}&=&-{1\over 2}\sqrt{-h}h^{\alpha\beta}
\partial_{\alpha}X^{\mu}\partial_{\beta}X^{\nu}
g_{\mu\nu}
+{1\over 2}\epsilon^{\alpha\beta}\partial_{\alpha}X^{\mu}
\partial_{\beta}X^{\nu}B_{\mu\nu}
\label{eq:Lb}\\
{\cal L}_{f}&=&-i\partial_{\alpha}X^{\mu}\overline{\Theta}
(\sqrt{-h}h^{\alpha\beta}-\epsilon^{\alpha\beta}\Gamma_{11})
\Gamma_{\mu}\tilde{{\cal D}}_{\beta}\Theta.
\label{eq:Lf}
\end{eqnarray}
Indices $\alpha$, $\beta$ denote the world-sheet directions
$\tau$ and $\sigma$. $h_{\alpha\beta}$ is the world-sheet
metric and $\epsilon^{\tau\sigma}=+1$. 
Spacetime indices are $\mu,\nu=0,1,\ldots,9$,
and the indices for the transverse directions
are $i,j=1,\ldots,8$.
We will use $m,n=0,1,\ldots,9$ for the tangent frame.
Fermion $\Theta$ is the 32-component Majorana spinor,
$g_{\mu\nu}$ and $B_{\mu\nu}$ are the background metric
and the NS-NS 2-form, respectively.  The
$\Gamma$-matrices $\Gamma_{m}$ in 10 dimensions satisfy
$\{\Gamma_{m},\Gamma_{n}\}=2\eta_{mn}$. 
Chirality matrix $\Gamma_{11}$ is given by 
$\Gamma_{11}=\Gamma_{0}\Gamma_{1}\cdots \Gamma_{9}$.
Also note $\Gamma_{\pm}=(\Gamma_{9}\pm
\Gamma_{0})/\sqrt{2}$ and $\Gamma^{\pm}=\Gamma_{\mp}$.
The conjugate spinor $\overline \Theta$ is defined by
$\overline \Theta=\Theta^{\dagger}\Gamma^{0}$. 
Curved-space $\Gamma$-matrices
$\Gamma_{\mu}$ are given by applying 
the vielbein on $\Gamma_{m}$.

The derivative $\tilde{{\cal D}}_{\alpha}$ is 
the `pull-back of the supercovariant derivative'
defined by 
\begin{equation}
\tilde{{\cal D}}_{\alpha}\Theta \equiv  \partial_{\alpha}\Theta 
+\partial_{\alpha}X^{\mu}\left( {1\over 4}\omega_{\mu}{}^{mn}
\Gamma_{mn} + \Omega_{\mu}\right) \Theta,
\end{equation}
where $\omega_{\mu}{}^{mn}$ is the spin connection, and
\begin{equation}
\Omega_{\mu}\equiv 
{1\over 8}\Gamma_{11}\Gamma^{\rho\sigma}F_{\mu\rho\sigma}
-{1\over 16}e^{\phi}(\Gamma_{11}\Gamma^{\rho\sigma}
F_{\rho\sigma}-{1\over 12}\Gamma^{\rho\sigma\lambda\tau}
F_{\rho\sigma\lambda\tau})\Gamma_{\mu}.
\end{equation}
Here, $F_{\mu\nu\rho}$ is the NS-NS 3-form field strength,
$F_{\mu\nu}$ and $F_{\mu\nu\rho\sigma}$ are the RR 2-form
and 4-form field strengths, respectively, and $\phi$ is
the dilaton background. The antisymmetrized product of 
$\Gamma$-matrices $\Gamma^{\mu_{1}\ldots \mu_{n}}$ is
defined with weight 1.

On the pp-wave background, the only non-vanishing
contribution from the spin connection is
$\omega_{+}{}^{mn}\Gamma_{mn}=
-\partial_{i}g_{++}\Gamma_{i}\Gamma_{-}$.
Since the NS-NS flux is absent and 
the non-vanishing components of the RR fluxes are
$F_{+i}$ and $F_{+ijk}$ in our case, 
$\Omega_{\mu}$ is written as
\[
\Omega_{+}=\tilde{\Omega}\Gamma_{-}\Gamma_{+},\quad
\Omega_{-}=0,\quad 
\Omega_{i}=\tilde{\Omega}\Gamma_{-}\Gamma_{i},
\]
where
\[
\tilde{\Omega}\equiv
{1\over 8}e^{\phi}\left(\Gamma_{11}\Gamma^{i}F_{+i}
-{1\over 6}\Gamma^{ijk}F_{+ijk}\right).
\]

We substitute the plane wave background (\ref{eq:planewaveenh})
for the $p=0$ enhan\c{c}on into the action, and
impose the light-cone gauge condition
\begin{equation}
\sqrt{-h}h^{\alpha\beta}=\eta^{\alpha\beta},\quad
X^{+}=\alpha' p^{+}\tau,\quad 
\Gamma_{-}\Theta=0.
\label{eq:LCgauge}
\end{equation}
The equations of motion for the transverse bosonic fields $X_{i}$
become
\begin{equation}
\partial_{\tau}^{2}X^{i}-\partial_{\sigma}^{2}X^{i}
+\tilde{m}^{2}_{i}X^{i}=0,
\end{equation}
where $\tilde{m}^{2}_{i}\equiv (\alpha' p^{+})^{2} m^{2}_{i}$.
$m^{2}_{i}$ are obtained in section 4.2.
There are four massless fields ($m_{w}=0$),
three massive fields with mass $m_{y}$ and 
one massive field with mass $m_{z}$.

The fermionic part of the Lagrangian in the 
light-cone gauge reads
\begin{equation}
{\cal L}_{f}=i(\alpha' p^{+})\left\{
\overline{\Theta}\Gamma_{+}\partial_{\tau}\Theta
+\overline{\Theta}\Gamma_{11}\Gamma_{+}\partial_{\sigma}\Theta
+(\alpha' p^{+})\overline{\Theta}\Gamma_{+}\tilde{\Omega}
\Gamma_{-}\Gamma_{+}\Theta\right\}.
\label{eq:LCfermi}
\end{equation}
We now decompose $\Gamma_{m}$ into $16\times 16$ blocks
following ref.~\cite{CvLuPoSt}:
\[
\Gamma_{+}= \pmatrix{0& \sqrt2\cr 0& 0}\,,\quad
\Gamma_{-}= \pmatrix{0& 0\cr \sqrt2 & 0}\,,\quad
\Gamma_{i} = \pmatrix{\gamma_{i} & 0\cr 0& -\gamma_{i}}\,,\quad 
\Gamma_{11}=\pmatrix{\gamma_{9}&0\cr 0& -\gamma_{9}}\
\]
where $\gamma_{i}$ are the SO(8) $\Gamma$-matrices which
satisfy $\{ \gamma_{i}, \gamma_{j}\}=\delta_{ij}$, and
$\gamma_{9}\equiv \gamma_{1}\cdots \gamma_{8}$.
In this representation, spinors which satisfy 
$\Gamma_{-}\Theta=0$ are of the form
\[ 
\Theta=\pmatrix{0\cr \theta}.
\]
We further decompose $\theta$ into a pair of
8 component spinors $\theta_{\pm}$
according to their SO(8) chiralities:
\begin{equation}
\theta=\theta_{+}+\theta_{-},\quad
\gamma_{9}\theta_{\pm}=\pm \theta_{\pm}.
\end{equation}
We can write (\ref{eq:LCfermi})  in the form 
\begin{equation}
{\cal L}_{f}=\sqrt{2}i(\alpha' p^{+})\left\{
\theta_{+}^{\dagger}(\partial_{\tau}+\partial_{\sigma})\theta_{+}
+\theta_{-}^{\dagger}(\partial_{\tau}-\partial_{\sigma})\theta_{-}
+2\theta_{+}^{\dagger}\tilde{\Omega}_{-}\theta_{-}
+2\theta_{-}^{\dagger}\tilde{\Omega}_{+}\theta_{+}\right\}
\end{equation}
where 
\[
\tilde{\Omega}_{\pm}\equiv {\alpha' p^{+}\over 8}e^{\phi}
\left( \mp\gamma_{i}F_{+i}+{1\over 6}\gamma_{ijk}F_{+ijk}\right).
\]

The equations of motion become
\begin{equation}
(\partial_{\tau}+\partial_{\sigma})\theta_{+}
+2\tilde{\Omega}_{-}\theta_{-}=0,\quad
(\partial_{\tau}-\partial_{\sigma})\theta_{-}
+2\tilde{\Omega}_{+}\theta_{+}=0.
\end{equation}
Bringing these equations into diagonalized 
second-order forms, we find that the masses
of the fermionic modes are given by
the eigenvalues of 
$-4\tilde{\Omega}_{+}\tilde{\Omega}_{-}$. 
Evaluating it with the plane wave fluxes 
(\ref{eq:fluxesrm}) for the fixed radius $r_{*-}$,
we obtain
\begin{eqnarray}
-4\tilde{\Omega}_{+}\tilde{\Omega}_{-}&=&
-{(\alpha' p^{+})^{2}\over 16}
\left\{-(e^{\phi}F_{+123})^{2}- (e^{\phi}F_{+4})^{2}+
2e^{2\phi}F_{+123}F_{+4}\gamma_{1234}\right\}\nonumber\\
&=&
{27\over 2}{(\alpha' p^{+})^{2}\over r^{2}_{*-}}
{(3x-3-h(x))(x-1-h(x))\over (x-5-h(x))^{3}(5x-1-h(x))^{3}}
\nonumber\\
&&\quad \times \{ 5x-1-h(x)\pm x(x-5-h(x))\}^{2}.
\label{eq:fermimassrm}
\end{eqnarray}
Here, indices $1,2,3$ denote the $y$-directions,
and $4$ denotes the $z$-direction. 
Signs $\pm$ in (\ref{eq:fermimassrm}) 
denote the eigenvalues of $\gamma_{1234}$.
In the decoupling limit, where the fluxes are given
by (\ref{eq:fluxes}), this reduces to
\begin{equation}
-4\tilde{\Omega}_{+}\tilde{\Omega}_{-}=
{(3\pm 1)^{2}\over 16}{(\alpha' p^{+})^{2}\over r^{2}_{*d}}.
\label{eq:fermimass}
\end{equation}
For the world-sheet fermions, we have four fields 
of the same mass $\tilde{m}^{2}_{f1}$ given by the plus signs of 
(\ref{eq:fermimassrm}) or (\ref{eq:fermimass}),
and four fields of the same mass $\tilde{m}^{2}_{f2}$ given by 
the minus signs of (\ref{eq:fermimassrm}) or (\ref{eq:fermimass}).

The frequency of the  oscillator at the $n$-th level
is given by 
$\omega_{n}=\sqrt{n^{2}+\tilde{m}^{2}}$, where 
$\tilde{m}^{2}$ is either the mass for the bosonic modes
$\tilde{m}^{2}_{i}$ or for the fermionic modes $\tilde{m}^{2}_{f1}$,
$\tilde{m}^{2}_{f2}$. We note here that the sum of $\omega^{2}_{n}$
for the bosonic oscillators is equal to that for 
the fermionic oscillators, at each $n$. 
Indeed, the following equality
(relation for the sum of $\omega_{0}^{2}$'s) holds
\[
4\times m^{2}_{w}+3\times m^{2}_{y}+m^{2}_{z}
=4\times m^{2}_{f1}+4\times m^{2}_{f2},
\]
as we see from (\ref{eq:massrm}) and (\ref{eq:fermimassrm}), 
or from (\ref{eq:massdec}) and (\ref{eq:fermimass}).
This property, which was also observed
for the plane waves for the Maldacena-N\'{u}\~{n}ez
and the Klebanov-Strassler solutions 
in ref.~\cite{GiPaSoSt}, guarantees that 
the string theory is finite.

Mode expansion can be performed in the standard way.
(See {\it e.g.} refs.~\cite{Me, BeMaNa}.)
We obtain the light-cone Hamiltonian as follows:
\begin{eqnarray}
H&=&H_{b0}+H_{b}+H_{f0}+H_{f}+E_{0}, 
\label{eq:hamiltonian}\\
H_{b0}&=&{1\over 2 p^{+}}\sum_{i=1}^{4}P_{i}^{2}+ 
{1\over 2 \alpha' p^{+}}\left(\sum_{i=5}^{7} 
\tilde{m}_{y}a^{i}_{0}{}^{\dagger}a^{i}_{0}
+\tilde{m}_{z}a^{z}_{0}{}^{\dagger}a^{z}_{0}
\right),\\
H_{b}&=&{1\over 2 \alpha' p^{+}}
\sum_{n=1}^{\infty}\left(
\sum_{i=5}^{7}\sqrt{n^{2}+\tilde{m}_{y}^{2}}
(a^{i}_{n}{}^{\dagger}a^{i}_{n}
+\tilde{a}^{i}_{n}{}^{\dagger}\tilde{a}^{i}_{n})
+\sqrt{n^{2}+\tilde{m}_{z}^{2}}
(a^{z}_{n}{}^{\dagger}a^{z}_{n}
+\tilde{a}^{z}_{n}{}^{\dagger}\tilde{a}^{z}_{n})
\right.
\nonumber\\
&&\left. \quad +\sum_{i=1}^{4} n
(a^{i}_{n}{}^{\dagger}a^{i}_{n}
+\tilde{a}^{i}_{n}{}^{\dagger}\tilde{a}^{i}_{n})
\right),\\
H_{f0}&=&{1\over 2 \alpha' p^{+}}
\sum_{A=1}^{2}\sum_{\rho=1}^{4} \tilde{m}_{f, A}
b^{A\dagger}_{0,\rho}b^{A}_{0,\rho}, \\
H_{f}&=&{1\over 2 \alpha' p^{+}}
\sum_{n=1}^{\infty}\sum_{A=1}^{2}\sum_{\rho=1}^{4}
\left(
\sqrt{n^{2}+\tilde{m}_{f, A}^{2}}
(b^{A\dagger}_{n,\rho}b^{A}_{n,\rho}
+\tilde{b}^{A\dagger}_{n,\rho}\tilde{b}^{A}_{n \rho})
\right).
\end{eqnarray}
Here $P_{i}$ $(i=1,\ldots, 4)$ is the center of mass
momentum in the $w$-directions. 
$a^{i}_{n}$ ($n\ge 0$; $i=1,\ldots, 8$)
and $\tilde{a}^{i}_{n}$ ($n\ge 1$;   $i=1,\ldots, 8$)
are the annihilation operators for 
the bosonic harmonic oscillators. 
$b^{A}_{n,\rho}$, $b^{A}_{n,\rho}$ $(A=1,2)$ 
are the fermionic harmonic oscillators, where
we have written the spinor indices
$\rho=1,\ldots, 4$ explicitly. 
These oscillators obey the standard commutation relations:
$[a^{i}_{n},a^{j\dagger}_{m}]=\delta^{ij}\delta_{nm}$,
$[\tilde{a}^{i}_{n},\tilde{a}^{j\dagger}_{m}]
=\delta^{ij}\delta_{nm}$,
$\{b^{A}_{n,\rho},b^{B\dagger}_{m,\sigma}\}=
\delta^{AB}\delta_{nm}\delta_{\rho\sigma}$,
$\{\tilde{b}^{A}_{n,\rho},
\tilde{b}^{B\dagger}_{m,\sigma}\}
=\delta^{AB}\delta_{nm}\delta_{\rho\sigma}$.
There should also be a vacuum energy $E_{0}$
in (\ref{eq:hamiltonian}). 
A physical state $|\psi\rangle$ 
is subject to the level matching
condition:
\[
\left( \sum_{n=1}^{\infty}\sum_{i=1}^{8}
n a_{n}^{i\dagger}a_{n}^{i}+
\sum_{n=1}^{\infty}\sum_{A=1}^{2}\sum_{\rho=1}^{4}
n b_{n,\rho}^{A\dagger}b_{n,\rho}^{A}\right)
|\psi\rangle
=
\left(
\sum_{n=1}^{\infty}\sum_{i=1}^{8}
n \tilde{a}_{n}^{i\dagger}\tilde{a}_{n}^{i}+
\sum_{n=1}^{\infty}\sum_{A=1}^{2}\sum_{\rho=1}^{4}
n \tilde{b}_{n,\rho}^{A\dagger}\tilde{b}_{n,\rho}^{A}\right)
|\psi\rangle.
\]

\section{Comment on the dual gauge theory}

The gauge theory which is dual to the string theory
on the decoupling limit of the $p=0$ enhan\c{c}on
geometry is the five dimensional super Yang-Mills theory
compactified on $K3$ with the volume $V$.
Firstly, we recall that the gauge coupling of the
five dimensional theory and that of the $K3$-compactified one
dimensional 
theory are given by
\begin{equation}
g_{YM,5}^{2}=(2\pi )^{2}g_{s}\alpha'{}^{1/2},\quad
g_{YM,1}^{2}=(2\pi )^{2}g_{s}\alpha'{}^{1/2}V^{-1},
\end{equation}
respectively.
For the energy scale larger than the inverse radius
of $K3$ ($E> V^{-1/4}$), the five dimensional gauge theory provides
a good description, and for the small energy scale 
$(E< V^{-1/4})$, the one dimensional gauge theory will take over.
The effective coupling at energy scale $E$
for the gauge theory in the large $N$ limit
will be given by the following dimensionless 
't Hooft couplings~\cite{JoPePo}:
\begin{equation}
g_{YM,5}^{2}N E^{1/2}\;\;(\mbox{for }E> V^{-1/4}),\qquad
g_{YM,1}^{2}N E^{-3/2}\;\;(\mbox{for }E< V^{-1/4}).
\label{eq:effcoup}
\end{equation}

Let us examine what is the parameter region of the
gauge theory to which our string-theory results
correspond. The string theory on the supergravity
background gives a good description when the
effective string coupling (determined by the dilaton v.e.v.)
and the curvature of the background measured in the
string unit are both small. Let us examine these 
conditions at the constant radius\footnote{%
It may be more appropriate to require that the 
above two conditions are satisfied throughout the
`near-shell' region $r_{e}\le r\le r_{4}$,
as assumed in the case of the holography for 
the D0-branes \cite{JeYo}. 
If we take this viewpoint, the condition for the smallness 
of the dilaton is replaced by $g_{s}\ll 1$.}
$r_{*d}=4^{1/3}\tilde{r}_{0}$.
We have 
\begin{eqnarray}
e^{\phi}\big|_{r_{*d}}
&=&g_{s}Z_{0}^{3/4}Z_{4}^{-{1/4}}\big|_{r_{*d}}
= 2^{-1}\cdot 3^{3/4} g_{s}
\left( {V_{*}\over V}\right)^{1/4}=
2^{-1}\cdot 3^{3/4} {\mu\over N},\\
\label{eq:dilatoncond}
R\big|_{r_{*d}}&=& Z_{0}^{-1/2}Z_{4}^{-1/2}r^{-2}\big|_{r_{*d}}
=4^{1/3}3^{-1/2}\pi^{-2/3}{\mu^{-{2/3}}\over \alpha'},
\end{eqnarray}
where $\mu\equiv g_{s}N
\left({V_{*}/ V}\right)^{1/4}$.
Thus, we need to set 
$\mu\gg 1$ fixed but large, and 
take the $N\to \infty$ limit, in order to have
the string coupling and the curvature both small.


Note that $\mu$
is the effective coupling (\ref{eq:effcoup})
at the energy scale $E=V^{-1/4}$. 
The effective coupling grows as we increase
the energy above $V^{-1/4}$, and also grows as we
decrease the energy below $V^{-1/4}$.
Thus, the condition $\mu\gg 1$ implies that
the gauge theory to which our string results
should correspond is strongly coupled
at every energy region\footnote{%
In ref.~\cite{JoPePo}, identification of weakly 
coupled descriptions at various energy scales was
attempted. It was assumed that $\mu\ll 1$, 
which is opposite to our requirement here.}.

We note a qualitative behavior of the string spectrum
as we vary the volume $V$ of the $K3$. 
The world-sheet mass is written as
\begin{equation}
m^{2}\sim {1\over r^{2}_{*d}}\sim 
\left({V_{*}\over V}g_{s}N\right)^{-{2/3}}{1\over \alpha'}
=\left(\left({V_{*}\over V}\right)^{3/ 4}\mu\right)^{-{2/ 3}}
{1\over \alpha'}.
\end{equation}
From this, we see that in the five dimensional limit (when $V/V_{*}\gg 1$), 
we have $m^{2}\alpha'\gg 1$,
and in the one dimensional limit (when $V/V_{*}\ll 1$), $m^{2}\alpha'\ll 1$.
This suggests that in the five dimensional limit, the energy  
of the low lying states are degenerate, since
$\omega_{n}
=\sqrt{n^{2}+(p^{+}\alpha' m)^{2}}\sim 
(p^{+}\alpha' m)+n^{2}/(p^{+}\alpha' m)$.
On the other hand, in the one dimensional limit, $m^{2}$ becomes
small, and the spectrum approaches that of 
the string in the flat space.

We note that there are natural candidates for the
gauge-theory operators which correspond to the string states
in the Penrose limit. The bosonic fields in the five dimensional
 super Yang-Mills
theory on the D4-branes are $A_{0}$, $A_{a}$, $\phi_{i}$,
where the indices $a(=1,\ldots,4)$ denote 
the spatial directions along the branes,
and the indices $i(=5,\ldots,9)$ denote the directions 
transverse to the branes.
Upon the K3 compactification, the one dimensional gauge field $A_{0}$ and
five scalars $\phi_{i}$ remain massless.
Following the idea of BMN \cite{BeMaNa},
the ground state of the string, which have angular momentum
in the $S^{4}$ in the transverse space, would be
identified as the trace of the product of $Z\equiv \phi_{8}+i\phi_{9}$,
${\rm Tr} (Z\cdots Z)$. Creation of the state by
the oscillators in the `$y$-directions' $a^{i}_{n}$ $(i=5,6,7)$
will correspond to the $\phi_{i}$ insertions in the
trace, oscillators
in the `$z$-direction' will be represented by replacing one $Z$ with
the covariant derivative $D_{0}Z$, and oscillators in the
`$w$-directions' $a^{a}_{n}$ $(a=1,\ldots,4)$ will be given
by replacing one $Z$ with $D_{a}Z$. There should be the 
ordering-dependent phase factors, to represent each string 
mode $n$, as in ref.~\cite{BeMaNa}.

We leave the explicit prediction of the 
correlation functions of the gauge theory for future study. 
In the case of ${\cal N}=4$ super Yang-Mills theory, 
only a subset of the Feynman diagrams contribute to 
the scaling dimensions of the operators studied by BMN~\cite{BeMaNa}.
Whether similar non-renormalization properties exist for the theory 
with half the maximal supersymmetry is a non-trivial question.
Indeed, the superstring spectrum for our case is not supersymmetric.
It would be necessary to add SUSY breaking term to the Lagrangian.
It is an interesting problem to reproduce 
the behavior of the spectrum from the gauge theory calculation.

\section{Conclusions}

In this paper, we have studied the Penrose limit along the null
geodesics with fixed radius both in general brane background and 
in enhan\c{c}on geometry.
In general background we have obtained the explicit formula for the plane
wave metric and discussed 
the condition that resulting string theories 
have tachyonic mass.
In enhan\c{c}on geometry associated with $D(p+4)$ ($p=0,1,2$) branes wrapped
on the $K3$ surface, we examined possible types of null geodesics with
fixed radius and obtained solvable superstring theories 
from the Penrose
limit of the $p=0$ enhan\c{c}on geometry.
This geometry is dual to the five dimensional super Yang-Mills theory 
wrapped on $K3$ with eight supercharges.

In order to formulate the holography for this background, and to
extract the information on the gauge theory from the string theory
on the null geodesic with constant radius,
further analysis is necessary.
It would be important to study the superstring theory 
on the Penrose limit along generic null geodesic, 
which has time-dependent masses, 
and regard the present result as a special limit. 
In particular, we note that the null geodesic with
constant radius found in this paper 
can be regarded as a limit of
a family of the `tunneling null geodesics' introduced in ref.~\cite{DoShYo}. 
These are the null geodesics in the space-time in which
$t$ and $\psi$ are simultaneously Wick rotated. Such a
geodesic passes the region where $f(r)\le 0$, in the 
notation of section 4.1. 
Authors of ref.~\cite{DoShYo} formulated the holography for 
$AdS_{5}\times S^{5}$, based on the string theory
along tunneling null geodesic which connect two points of the
base space of the gauge theory living on the 
boundary of the  background in the decoupling limit.
The null geodesic with constant radius found for our background
corresponds to the limiting case where the two points on the boundary are 
infinitely separated. This suggests that the string 
on this special geodesic represents the gauge theory
in the IR limit. By clarifying the interpretation
of the null geodesic with constant radius, it might be
possible to have an explanation for the fact that
such a geodesic exist only for $p=0$ case from the 
gauge-theory point of view.

Another direction for future study would be
to apply the analysis of the present paper
on the null geodesics to other geometries. 
A possibly interesting background is the Pilch-Warner solution \cite{PiWa2} 
corresponding to the flow from ${\cal N}=4$ to ${\cal N}=2$ supersymmetric 
Yang-Mills theory. The massless field content of the above 
${\cal N}=2$ theory is the one for the pure ${\cal N}=2$ 
super Yang-Mills theory, which is the same as the dual theory for the enhan\c{c}on geometry given by the D7-brane wrapped on $K3$.
In refs.~\cite{BuPePo}, it was shown that the enhan\c{c}on
mechanism also takes place for this ${\cal N}=2$ Pilch-Warner solution.
It might be the case that the structure of the holographic RG
flow has some similarities with the ones for our enhan\c{c}on geometries.

Recently, the authors of ref.~\cite{GiPaSoSt} proposed an
interpretation of the string spectrum
at fixed radius on the Maldacena-N\'{u}\~{n}ez 
and the Klebanov-Strassler solutions. 
It is argued that the string states correspond to 
stable hadrons which are described by composite operators
made from massive fields. 
It is an interesting question whether 
a similar interpretation can be applied to our case.

\vspace{1cm}

\noindent{{\bf Acknowledgements}}

YS would like to thank M.~Egoshi and T.~Yoneya 
for valuable discussions.
We are also grateful to H.~Fuji for useful discussion.

\section*{Appendix: Null geodesics with constant
radius for various brane solutions}
\renewcommand{\thesubsection}{A.\arabic{subsection}}
\renewcommand{\theequation}{A.\arabic{equation}}
\setcounter{equation}{0}

As an application of the general procedure described 
in section 3, we analyze the D$p$-brane, $(p,q)$ fivebrane
$(p,q)$ string solutions, which have been studied
in the previous paper~\cite{FuItSe}.
We examine the conditions for the 
existence of the null geodesic with fixed radius
for each brane solution. When such a geodesic
exist, we take the Penrose limit. We note here
again that we have changed the sign in the definition
of the $m^{2}_{i}$ from ref.~\cite{FuItSe}.

We investigate the D$p$-branes in appendix A.1, 
the $(p,q)$ fivebranes including the NS5-branes
in appendix A.2, the $(p,q)$ strings including 
the fundamental strings in appendix A.3.
We will see that  a stable geodesic for which
all $m^{2}_{i}$ are non-negative,
is possible only for the fivebrane solutions
in the near-horizon limit.

\subsection{D$p$-branes}
The D$p$-brane solution in the string frame is given by
\begin{eqnarray}
&&ds^{2}=H^{-1/2}(-dt^{2}+dx^{2}_{a})
+H^{1/2}(dr^{2}+d\Omega^{2}_{8-p}),\nonumber\\
&&H=1+{Q_{p}\over r^{7-p}}.
\label{eq:Dpmetric}
\end{eqnarray}
We can apply the formulas of section 3 by 
substituting $A^{2}=H^{-1/2}$, $B^{2}=H^{1/2}$,
and $D=10$. We study the D$p$-branes with $0\le p \le 6$.
To study the behavior of the null geodesics on this
background, it is convenient to define $f(r)$ 
which is related to $F(r)$ in (\ref{eq:defFr}) by
\begin{equation}
f(r)\equiv r^{5-p}\ell^{2}H^{1/2} F(r).
\end{equation}
The conditions $\dot{r}=\ddot{r}=0$ 
for the null geodesic with
fixed radius are equivalent to
\begin{eqnarray}
&&f(r)=r^{7-p}-\ell^{2}r^{5-p}+Q_{p}=0,
\label{eq:Dpfr}\\
&&f'(r)=(7-p)r^{6-p}-(5-p)\ell^{2}r^{5-p}=0.
\label{eq:Dpfrp}
\end{eqnarray}
Note that the sign of $m_{z}^{2}$ 
is the same as that of $f(r)$.

In the near-horizon limit, the metric is
given by replacing $H$ in (\ref{eq:Dpmetric}) with 
\begin{equation}
H={Q_{p}\over r^{7-p}}.
\end{equation}
In this case, the conditions for 
$\dot{r}=\ddot{r}=0$ become
\begin{eqnarray}
&&f(r)=-\ell^{2}r^{5-p}+Q_{p}=0,
\label{eq:nhDpfr}\\
&&f'(r)=-(5-p)\ell^{2}r^{5-p}=0.
\label{eq:nhDpfrp}
\end{eqnarray}

We shall examine whether there are solutions to
(\ref{eq:Dpfr}) and (\ref{eq:Dpfrp}),
or (\ref{eq:nhDpfr}) and (\ref{eq:nhDpfrp})
for each brane background.

For the D$p$-branes with $p\le 4$, we have the solution 
$r=r_{*}$, $\ell=\ell_{*}$ of (\ref{eq:Dpfr})
and (\ref{eq:Dpfrp}) 
\begin{eqnarray}
r_{*}&=&\left( {5-p\over 2}Q_{p} \right)^{1\over 7-p},\nonumber\\
\ell_{*}^{2}&=&(7-p)(5-p)^{-{5-p\over 7-p}}2^{-2\over 7-p}
Q^{2\over 7-p}.
\end{eqnarray}
Penrose limit along the null geodesic 
fixed at the radius $r_{*}$ is obtained from 
the formula in section 3.
We get the plane wave metric (\ref{eq:planewave}) 
with
\begin{equation}
m_{x}=0,\quad
m_{y}={1\over r^{2}_{*}},\quad 
m_{z}=-{5-p\over r_{*}^{2}}.
\end{equation}
Since we have $m^{2}_{z}<0$, string theory has
tachyonic mass on this plane wave.

In the near-horizon limit, there is no solution to
(\ref{eq:nhDpfr}) and (\ref{eq:nhDpfrp}). Indeed,
this is clear from the fact that the fixed
radius $r_{*}$ is of the same order as the boundary
of the near-horizon region $r\sim Q^{1\over 7-p}$.

For $p=5$, (\ref{eq:Dpfr}) and
(\ref{eq:Dpfrp}) is satisfied by
$r_{*}=0$ and $\ell_{*}^{2}=Q_{5}$. Taking the 
Penrose limit along this geodesic, we get
$m_{y}^{2}$, $m_{z}^{2}$ which are divergent.
On the other hand, when we consider the near-horizon 
geometry, (\ref{eq:nhDpfr}) and (\ref{eq:nhDpfrp}) 
are satisfied by an arbitrary $r_{*}$ if  
$\ell^{2}_{*}=Q_{5}$ is satisfied.
The Penrose limit is given by
\begin{eqnarray}
&&m^{2}_{x}=m^{2}_{z}=0,\\
&&m^{2}_{y}={1\over r_{*}^{2}}.
\end{eqnarray}
In this case, we have $m^{2}_{z}=0$, and there
is no tachyonic mass on the world-sheet of the 
string on this background.
The Penrose limit of fivebrane solutions along the
null geodesic with constant radius was obtained
by Oz and Sakai \cite{OzSa}. 

For $p=6$, the conditions for the constant radius
(\ref{eq:Dpfr}) and (\ref{eq:Dpfrp}), or 
(\ref{eq:nhDpfr}) and (\ref{eq:nhDpfrp}) do not
have solutions.

\subsection{$(p,q)$ fivebranes}
The $(p,q)$ fivebrane metric in the string frame is
\begin{eqnarray}
&&ds^2 =h^{-1/2}
(-dt^2+ dx_a^2 + \tilde{H} 
(dr^2+ r^2 d\Omega_{3}^2)), \nonumber \\
&&\tilde{H}=1+{\tilde{Q}_5\over  r^2},\quad
h^{-1}=\sin^2\gamma \tilde{H}^{-1} +\cos^2\gamma.
\label{eq:pq5}
\end{eqnarray}
where $\tan \gamma=q/p$. 
This includes the NS5-branes ($\cos\gamma=1$)
and the D5-branes ($\cos\gamma=0$) discussed above.
The conditions for the fixed radius $\dot{r}=\ddot{r}=0$ 
are equivalent to $f(r)=f'(r)=0$, where
\begin{equation}
f(r)\equiv h^{1/2}\ell^{2} F(r)
=r^{2}-\ell^{2}+\tilde{Q}_{5}.
\label{eq:pq5fr}
\end{equation}
The conditions are of the similar form
as for the D5-branes, i.e. the $p=5$ case of (\ref{eq:Dpfr}).
(\ref{eq:pq5fr}) can be satisfied by 
$r_{*}=0$ and $\ell_{*}^{2}=\tilde{Q}_{5}$, but
$m_{y}^{2}$, $m_{z}^{2}$ are divergent.

The $(p,q)$ fivebrane metric in the near-horizon
limit is given by
replacing $\tilde{H}$ in (\ref{eq:pq5}) with
\begin{equation}
\tilde{H}={\tilde{Q}_{5}\over r^{2}}.
\end{equation}
The conditions for the constant radius is satisfied
by arbitrary $r_{*}$, if we have $\ell^{2}_{*}=\tilde{Q}_{5}$.
In the Penrose limit, we have
\begin{eqnarray}
&&m^{2}_{x}=m^{2}_{z}=0,\\
&&m^{2}_{y}={1\over \tilde{Q}_{5}\cos^{2}\gamma
+\tilde{r}_{*}^{2}\sin^{2}\gamma},
\end{eqnarray}
which is consistent with the result
given in ref.~\cite{OzSa}. 
As in the case of the D5-branes, we have
$m^{2}_{z}=0$ which is non-negative,
for the $(p,q)$ fivebranes in the near-horizon limit.

\subsection{$(p,q)$ strings}
The $(p,q)$ string metric in the string frame 
is given by
\begin{eqnarray}
&&ds^2=h^{-1/2}(
\tilde{H}^{-1}(-dt^2+dx^2) + dr^2+ r^2 d\Omega_{7}^2),\nonumber\\
&&\tilde{H}=1+{\tilde{Q}_1\over  r^6},\quad
h^{-1}=\sin^2\gamma \tilde{H} +\cos^2\gamma. 
\label{eq:pq1}
\end{eqnarray}
where $\tan \gamma=q/p$. 
We obtain the fundamental strings by setting
$\cos\gamma= 1$, and the D1-branes discussed above
by setting $\cos\gamma= 0$. 
The conditions $\dot{r}=\ddot{r}=0$ are equivalent 
to $f(r)=f'(r)=0$, where 
\begin{equation}
f(r)\equiv h^{1/2}\ell^{2}r^{4}\tilde{H} F(r)\nonumber\\
=r^{6}-\ell^{2}r^{4}+\tilde{Q}_{1}.
\label{eq:pq1fr}
\end{equation}
Since (\ref{eq:pq1fr}) is of the same form as 
the $p=1$ case of (\ref{eq:Dpfr}), the conditions for
the constant radius are the same the ones for the D1-branes:
\begin{eqnarray}
r_{*}&=&(2\tilde{Q}_{1})^{1\over 6}\nonumber\\
\ell_{*}^{2}&=&3\cdot 2^{-{2\over 3}}\tilde{Q}_{1}^{1\over 3}.
\end{eqnarray}
Taking the Penrose limit, we obtain
\begin{eqnarray}
m^{2}_{x}&=&0,\\
m^{2}_{y}&=&{3\over 3-\cos^{2}\gamma}{1\over r^{2}_{*}},\\
m^{2}_{z}&=&-{12\over 3-\cos^{2}\gamma}{1\over r^{2}_{*}}.
\end{eqnarray}
We have negative mass-squared $m^{2}_{z}<0$.

The near-horizon limit of the
$(p,q)$ string metric is given by
replacing $\tilde{H}$ in (\ref{eq:pq1}) with
\begin{equation}
\tilde{H}={Q_{1}\over r^{6}}.
\end{equation}
In this case, there is no null geodesic with
constant radius.

\end{document}